\documentclass[prd,aps,showpacs,nofootinbib,floatfix,onecolumn,superscriptaddress]{revtex4-2}

\usepackage{slashed}
\usepackage{float}
\usepackage{amsmath,graphicx,color,epsfig,ulem}
\usepackage{bm}
\usepackage{cancel}
\definecolor{purple}{rgb}{0.5,0,0.5}
\usepackage[colorlinks=true, pdfstartview=FitV, linkcolor=purple, citecolor=blue, urlcolor=blue]{hyperref}

\newcommand{\beq}{\begin{equation}}
\newcommand{\eeq}{\end{equation}}

\begin{document}

\title{Photon Gas Thermodynamics in Doubly Special Relativity at the Planck Scale}

\author{Qi Xiong}
\affiliation{Department of Physics, Shanghai Normal University, Shanghai 200234, People's Republic of China}
\affiliation{National Astronomical Observatories, Chinese Academy of Sciences, 20A Datun Road, Beijing 100101, People's Republic of China}

\author{Xinyi Yang}
\affiliation{Department of Physics, Shanghai Normal University, Shanghai 200234, People's Republic of China}

\author{Guifeng Su}
\affiliation{Department of Physics, Shanghai Normal University, Shanghai 200234, People's Republic of China}

\author{Yi Zhang}
\email[Corresponding author: ]{yizhang@shnu.edu.cn}
\affiliation{Department of Physics, Shanghai Normal University, Shanghai 200234, People's Republic of China}

\begin{abstract}
 We investigate the thermodynamics of a photon gas within the Magueijo-Smolin formulation of doubly special 
 relativity, a framework that augments special relativity with an observer-independent energy scale of the 
 order of the Planck energy. In this realization the photon dispersion relation is undeformed, so that the 
 deformation enters the statistical calculation only through the observer-independent upper bound $E_{\rm P}$ 
 on the single-particle energy. We derive the logarithmic grand partition function, and the complete set of 
 thermodynamic quantities for the photon gas, including the Helmholtz free energy, internal energy, entropy, 
 pressure, and heat capacity, and perform their numerical evaluation. Our results smoothly reduce to the 
 conventional special relativistic expressions in the limit where the invariant energy scale tends to infinity. 
 For temperatures approaching the Planck scale, the finite cutoff induces a systematic suppression of 
 the absolute value of Helmholtz free energy, internal energy, entropy, pressure, and heat capacity, relative 
 to their standard special-relativistic counterparts. We identify the boundary of the regime in which this 
 description remains self-consistent.
\end{abstract}


\maketitle

\section{Introduction}

Although a complete and self-consistent theory of quantum gravity remains elusive, general considerations 
rooted in established physics suggest that novel effects should emerge near the Planck scale. Any viable 
candidate theory is expected to reproduce special relativity (SR) in regimes where gravitational effects 
are negligible and the relevant energy scales lie far below the Planck energy, 
$E_{\rm{P}} = \sqrt{\hbar c^5/G} \sim 10^{19}$ GeV, 
or equivalently, the Planck length, $\ell_{\rm{P}} = \sqrt{\hbar G/c^3} \sim 10^{-35}$ m. 
Beyond these scales, the classical description of spacetime breaks down, and quantum gravitational effects 
are expected to become significant. This naturally raises a fundamental question: with respect to which 
inertial frame is the Planck scale defined? For instance, if a microscopic length associated with quantum 
spacetime structure is measured as the Planck length $\ell_{\rm P}$ in one frame, Lorentz contraction 
would imply a different value in another frame, which is problematic given that $\ell_{\rm P}$ is 
constructed from the fundamental constants $\hbar$, $G$, and $c$, and is thus supposed to be 
observer-independent.

To address this tension, several nonlinear extensions of SR have been proposed~\cite{Ameli2001a,
Ameli2001b,Glik2001,Ameli2002a,Ameli2002b,Glik2002a,Glik2002b,Mague2002,Mague2003,Glik2003,Mague2004}. 
In standard SR, physical laws rest on the relativity principle and the invariance of the speed of light, 
with inertial-frame transformations given by linear Lorentz transformations. In contrast, nonlinear 
relativity preserves covariance while allowing nonlinear transformation laws between reference frames. 
A particularly prominent realization is deformed, or doubly special relativity (DSR), which retains the 
invariant speed of light and introduces a second observer-independent scale, typically identified with 
the Planck length or energy. The first model of this kind was proposed by 
Amelino-Camelia~\cite{Ameli2001b,Ameli2002a}. Subsequently, a second type of DSR model was introduced 
by Amelino-Camelia~\cite{Ameli2002b} and independently by Magueijo and Smolin~\cite{Mague2002,Mague2003} 
(the latter one is referred to as the MS model in the present study). For recent introductions and 
reviews about DSR, see, e.g., Refs.~\cite{Glik2005,Ameli2010a} and references therein. In MS models, 
there are two invariant quantities: the speed of light $c$ and an upper energy scale $E_{\rm P}$, both 
postulated to be observer-independent. To consistently incorporate this second invariant alongside the 
other principles of SR, the standard dispersion relation $E^2 - p^2 = m^2$ must be modified to 
$E^2 - p^2 = m^2 \left(1 - \frac{E}{E_{\rm P}}\right)^2$ (in $c=1$ units), where $E$ and $p$ denote 
the energy and the magnitude of the three-momentum of the particle, respectively. 
The parameter $m$ is the invariant mass parameter: it is left unchanged by a DSR transformation and 
is therefore not the rest mass of the particle in the SR sense. 
Consequently, the integration over energy-momentum space is also modified~\cite{Ameli2000,Ameli2009}.

These modifications have led to further investigations and various predictions, including cosmic ray 
physics~\cite{Ameli2001a,Ameli2002b,Gamb1999,Ellis2000}, relativistic thermodynamics~\cite{Chan2012}, 
black hole thermodynamics~\cite{Sales2009,Ameli2006}, cosmology~\cite{Alex2003}, compact 
stars~\cite{Greg2009,Berto2010}, neutrino propagation~\cite{Alfaro1999}, dark energy~\cite{Mersini2001}, 
and relativistic quantum mechanics~\cite{Menc2013,Jafa2024,Guven2025}, to name but a few. DSR is also 
closely connected to noncommutative spacetime structures~\cite{Glik2002a,Glik2002b,Glik2003,Ghosh2007}. 
In particular, the so-called \(\kappa\)-Minkowski spacetime provides a concrete realization in which 
the canonical Poisson brackets among phase space variables are modified. Correspondingly, the modified 
dispersion relation becomes invariant under nonlinear $\kappa$-Lorentz transformations, rather than 
under the standard linear Lorentz transformations~\cite{Bruno2001,Ghosh2007}, while the Lorentz 
algebra itself remains intact. For recent reviews of quantum spacetime phenomenology and new 
perspectives within DSR framework, see Refs.~\cite{Ameli2010a,Ameli2010b,Ameli2013,Carm2019,Carm2023}.

Motivated by these developments, in the present paper we adopt the MS framework as the basis for our 
investigation of photon gas thermodynamics in the Planck-scale regime. Although the impact of modified 
dispersion relations (MDR) on photon gas thermodynamics has been examined in several existing 
studies~\cite{Cama2007,Das2010,Zhang2011} (to the best of our knowledge), the necessity of the present 
investigation will be clarified below.

In Ref.~\cite{Cama2007}, the authors introduced a modified dispersion relation as a fundamental 
assumption for photon dynamics and analyzed its effects on the thermodynamics of a photon gas. As a 
result, the breakdown of Lorentz symmetry entails an increase in the number of microstates and, 
consequently, an enhancement of entropy relative to the case in which Lorentz symmetry remains unbroken. 
It should be noted that, in this study, the speed of light $c = dE/dp$ is variable and energy dependent. 
Although the model admits an upper bound on momentum, it does not impose a finite upper bound on photon 
energy.

In contrast, the MS model is characterized by two invariants: the speed of light $c$ and an upper energy 
scale $E_{\rm P}$. This double invariance leads to the specific form of the aforementioned MDR. 
Furthermore, for massless photons ($m = 0$), the dispersion relation remains identical to that of standard 
SR~\cite{Mague2002,Mague2003,Ghosh2007}. This MDR serves as the fundamental assumption and starting point 
adopted in Ref.~\cite{Das2010}. However, the treatment in that work employed the Maxwell-Boltzmann (MB) 
distribution in computing the photon gas partition function, which is inherently inappropriate for 
describing photon gas thermodynamics. The correct equilibrium description of a photon gas requires the 
grand canonical ensemble with Bose-Einstein (BE) statistics and vanishing chemical potential 
($\mu = 0$)~\cite{KH1987}. Consequently, the MB distribution fails to recover the standard SR photon gas 
thermodynamics in the appropriate limit.

Photon gas thermodynamics within the MS framework was also considered in Ref.~\cite{Zhang2011}. 
The present study differs from that work in the choice of dispersion relation (undeformed photon 
vs.\ $p=E/(1+\lambda E)$) and in the status of the momentum-space measure (an explicit modeling 
choice with $\theta=0$ vs.\ derived from the composition law). As emphasized in Secs.~\ref{sec:DSR} 
and \ref{sec:PF}, we treat the flat counting measure ($\theta=0$) as an explicit modeling choice, 
and we enforce the hard cutoff $E<E_{\rm P}$ as a covariant statement valid in every inertial 
frame. We also identify which of our conclusions -- the exponentially suppressed deviations at low 
temperatures and the cutoff artifacts at $T\gtrsim E_{\rm P}$ -- depend on these choices.

A further, and in our view more fundamental, concern is the kinematic input that any statistical-mechanical 
calculation presupposes. Because the deformed variables are related to the standard ones by a nonlinear map, 
the invariant volume element of momentum space is not the flat one, and the literature on DSR thermodynamics 
is not uniform in its choice of measure. As we shall show in Secs.~\ref{sec:DSR} and \ref{sec:PF}, this is 
not a technicality. We therefore treat the measure as an explicit ingredient of the model rather than as an 
implicit assumption, and we identify which of our conclusions depend on it. This issue is, in our view, more 
consequential than the numerical details of the resulting equations of state, and the paper is organized 
accordingly.

The remainder of this paper is organized as follows. In Sec.~\ref{sec:DSR}, we briefly review the DSR 
framework and the fundamental modified dispersion relation in the MS model, and we examine the kinematics 
of the deformed boost and the associated phase-space measures; Sec.~\ref{sec:DSR} also contains a 
comparison with the Amelino-Camelia realization. In Sec.~\ref{sec:PF}, we state the statistical-mechanical 
assumptions on which our treatment rests, introduce the phase-space measure and derive the logarithmic 
grand partition function for a photon gas subject to the Planck-scale upper bound on single-particle 
energies. In Sec.~\ref{sec:thermo}, based on this logarithmic partition 
function, we derive and numerically evaluate the associated thermodynamic quantities, discuss their 
limiting behaviors, and establish the range of validity of the description. As expected, the standard SR 
results are recovered in the low energy limit. We summarize our conclusions in Sec.~\ref{sec:summary}.

\section{The MDR and measure in the MS Model}
\label{sec:DSR}

\subsection{The MDR in Magueijo-Smolin model}

We begin with the well-known MS model~\cite{Mague2002,Mague2003,Mague2004}. 
This formulation preserves the relativity principle while elevating the Planck energy to an invariant 
scale for all inertial observers. 

Within a generic isotropic DSR framework, the MDR takes the general form
\beq 
\label{eq:mdr}
 E^2 f_1^2(E,E_{\rm P})-p^2 f_2^2(E,E_{\rm P}) = m^2 ,
\end{equation}
where $E$, $p$, and $m$ denote the particle energy, the magnitude of its three-momentum, and its 
mass, respectively. The functions $f_1(E,E_{\rm P})$ and $f_2(E,E_{\rm P})$ are model-dependent 
deformation functions, and $E_{\rm P}$ represents the observer-independent energy scale, 
conventionally taken to be of the order of the Planck energy.

By taking $f_1^2= f_2^2 = \left(1-\frac{E}{E_{\rm P}} \right)^{-2}$ in the MS model, 
the above MDR reduces to the specific form (in $c=1$ units).
\beq
\label{eq:ms_mdr}
  E^2-p^2=m^2\left(1-\frac{E}{E_{\rm P}}\right)^2 .
\eeq

It should be borne in mind, however, that this choice is model-dependent, and different DSR 
realizations may introduce distinct deformations of the relativistic dispersion relation; 
the physically viable form must ultimately be constrained by observational or experimental 
data~\cite{Ameli2002a,Ameli2010a}.

Owing to the MDR in Eq.~\eqref{eq:ms_mdr}, it has been shown that such a modification does 
not necessarily imply an energy-dependent speed of light~\cite{Hoss2006}. Thus, in the MS model, 
the dispersion relation for photons remains identical to that of standard SR, and the speed of 
light $c$ remains an invariant~\cite{Mague2002,Mague2003}. Nevertheless, the photon energy is 
subject to a finite upper bound set by the Planck scale $E_{\rm P}$, which is invariant under 
the deformed Lorentz transformation law for energy~\cite{Bruno2001,Ghosh2007}.

This physical picture immediately leads to a scenario that differs from that of 
Ref.~\cite{Cama2007}. In that study, the explicit breaking of Lorentz symmetry results in an 
increase in the number of microstates and hence in the entropy relative to the standard SR. 
In the present framework, by contrast, the ordinary \textit{linear} Lorentz symmetry is replaced 
by a nonlinear deformed realization, while the Lorentz \textit{algebra} itself remains undeformed, 
and it is the phase space that is constrained by the finite upper bound on energy. Consequently, 
one naturally expects a reduction in the number of microstates and, correspondingly, a decrease 
in entropy compared to the standard SR case, as we will see in Sec.~\ref{sec:thermo}.

\subsection{The momentum space measure}

Before turning to the statistical mechanics, however, we must specify the kinematic input that a 
continuum integration over momentum space presupposes. As we shall see, this point deserves care, 
because the choice of measure is not a technical detail.

Let $(e,\vec\pi)$ denote the standard special-relativistic energy-momentum variables, on which the 
ordinary Lorentz transformations act linearly, and $(E,\vec p)$ the corresponding MS variables. 
In the MS realization the two sets are related by the invertible nonlinear map~\cite{Mague2002,Mague2003}
\beq
\label{eq:map}
 E=\frac{e}{1+e/E_{\rm P}} , \qquad \vec p=\frac{\vec\pi}{1+e/E_{\rm P}} ~,
\eeq
and the inverse map is
\beq
 e=\frac{E}{1-E/E_{\rm P}} , \qquad \vec\pi=\frac{\vec p}{1-E/E_{\rm P}} ~.
\eeq
Eliminating $e$ and $\vec\pi$ in favour of $E$ and $\vec p$ in the invariant $e^2-\vec\pi^2=m^2$ 
reproduces Eq.~\eqref{eq:ms_mdr} exactly, while the deformed boosts act \textit{linearly} on 
$(e,\vec\pi)$. Because the map \eqref{eq:map} is nonlinear, three distinct objects, which are under 
the single name ``the measure'' routinely conflated in the literature, must be separated.

(i) \textit{The invariant measure on the four-dimensional momentum space.} Taking the Jacobian of 
Eq.~\eqref{eq:map}, one finds
\beq
\label{eq:inv4d}
d^4\Pi = \left(1-\frac{E}{E_{\rm P}}\right)^{-5} d^4 p ,
\eeq
which is the volume element of a doubly special (de Sitter) momentum space written in MS coordinates and 
coincides with the measure obtained for the same model in Ref.~\cite{Gubito2016}.

(ii) \textit{The invariant measure on the mass shell}, i.e., the image of the SR invariant measure 
$d^3\pi/(2e)$:
\beq
\label{eq:covmeasure}
 d\Gamma = \frac{d^3 p}{2\left(1-|\vec p|/E_{\rm P}\right)^3 |\vec p|} \qquad (m=0 ,\ c=1) ,
\eeq
which differs from $d^3p/(2E)$ by the factor $(1-E/E_{\rm P})^{-3}$ and is singular as 
$|\vec p|\to E_{\rm P}$~\cite{Rembie2003}.

(iii) \textit{The counting (mode) measure} $d^3p/(2\pi\hbar)^3$ per mode, i.e., the volume element of the 
deformed phase space. This is the quantity that enters the density of states $g(E)$, and therefore the 
one implicitly used in Eq.~\eqref{eq:lnXi_dsr1} below.

The four-momentum volume integral, the on-shell momentum integral, and the six-dimensional phase-space 
mode count serve distinct purposes and cannot be freely interchanged, even within SR, although a standard 
equivalence holds for flat Minkowski momentum space equipped with the conventional dispersion relation. 
This distinction carries direct implications for our statistical treatment. If the on-shell invariant 
measure, Eq.~\eqref{eq:covmeasure}, is adopted as the density of states, the single-particle partition 
function acquires a factor $\left(1-E/E_{\rm P}\right)^{-3}$, which diverges at the upper kinematic bound; 
accordingly, $\ln \Xi$, $U$, and $S$ are ill-defined. The four-dimensional invariant measure, 
Eq.~\eqref{eq:inv4d}, leads to the same difficulty. This is the core observation reported in Ref.~\cite{Rembie2003}.

The divergence depends on both the integration measure and the Hamiltonian entering the thermal weight. 
In the present MS prescription, the physical photon energy is bounded by $E_{\rm P}$, so the Boltzmann 
factor approaches the nonzero value $e^{-\beta E_{\rm P}}$ at the endpoint. It therefore cannot suppress 
the non-integrable endpoint singularity of the proposed weight. Rewriting the same integral in terms of 
the auxiliary energy $e = E/(1 - E/E_{\rm P})$ does not remove this divergence: the thermal factor remains 
$\exp [-\beta e/(1 + e/E_{\rm P})]$. Using $e^{-\beta e}$ instead would amount to choosing the auxiliary 
energy as the thermal Hamiltonian and would define a different statistical model. Other DSR realizations 
require a separate convergence analysis based on their measures and thermal Hamiltonians. 
Models based on de Sitter momentum space provide examples with different kinematic and statistical 
inputs~\cite{Gorji2016}; their convergence properties should not be attributed to a change of coordinates 
in the present MS calculation.

The flat measure is sometimes justified in the DSR literature by asserting that the phase-space element 
$d^3x\,d^3p$ is invariant under the deformed boosts, and hence that the density of states is unmodified. 
This claim unfortunately does \textit{not} hold. The exact Jacobian of the MS boost on momenta, obtained 
from Eq.~\eqref{eq:map} together with the linear action of the boost on $\vec\pi$, is
\beq
\label{eq:jacboost}
\det\frac{\partial \vec p'}{\partial \vec p}
= \frac{E'}{E}\,\alpha^{-3}
= \gamma\left(1-\frac{vp_x}{E}\right)\alpha^{-4} ,
\qquad
\alpha \equiv 1+\frac{(\gamma-1)E-\gamma vp_x}{E_{\rm P}} ,
\eeq
whereas Ref.~\cite{Das2010} quotes $\gamma\left(1-vp_x/E\right)\alpha^{-3}$. The missing power --- a factor 
$\alpha$ --- is exactly the variation $d\alpha$ that is discarded there as ``a dynamical effect'', yet it is 
this very term that carries the deformation of the measure, so that this approximation effectively assumes 
the result rather than deriving it. 
Independently of this, the momentum-space Jacobian alone is insufficient to establish invariance of the full 
phase-space measure: even in the undeformed limit $\alpha\to 1$, Eq.~\eqref{eq:jacboost} does not reduce to 
unity but to the familiar flux 
factor $\gamma(1-vp_x/E)$ of relativistic kinetic theory, reflecting the well-known fact that 
$d^3x\,d^3p$ is not a Lorentz invariant even before deformation, the invariant objects being $d^3p/E$ and 
$d^4x$. 

The object that counts microstates is the symplectic (Liouville) volume of the phase space. In the 
$\kappa$-Minkowski chart the spatial brackets of the deformed phase space are exactly canonical, 
$\{X_i,P_j\}=\delta_{ij}$~\cite{Blaut2004}, and in that chart the state count is $(2\pi\hbar)^{-3}\int d^3P$, 
i.e.\ flat, the whole deformation residing in the dispersion relation and in the boundedness of the momentum 
domain. Translating to the MS momentum variables generically introduces a factor $(1-E/E_{\rm P})^{-\theta}$, 
and the value of $\theta$ is not fixed by the MS kinematics alone, which is why the integration measure is 
treated as an independent choice in Ref.~\cite{Gubito2016}. Ref.~\cite{Zhang2011} derives the measure from 
the composition law, whereas the present work and Gubitosi-Magueijo treat it as a model choice --- which is 
precisely a manifestation of the fact that the measure issue remains unsettled in the literature. We adopt 
$\theta=0$: the chart in which the counting measure is flat and the deformation is carried entirely by the 
domain $0<E<E_{\rm P}$. 

\subsection{Comparison with the Amelino-Camelia realization}
\label{sec:AC}

It is instructive to contrast the situation described above with that in the Amelino-Camelia (AC) 
realization~\cite{Ameli2001b,Ameli2002a}, in which the second invariant is constructed differently. There 
the dispersion relation may be written as~\cite{Ameli2002b,Jafa2024,Jafa2025}
\beq
\label{eq:ACdisp}
2\kappa^2\left[\cosh\frac{E}{\kappa}-\cosh\frac{m}{\kappa}\right]=p^2 e^{E/\kappa} ,
\eeq
where $\kappa$ denotes the invariant energy scale. 
For $m=0$ this gives $p=2\kappa\,e^{-E/2\kappa}\sinh\!\big(E/2\kappa\big)$, i.e.\ $E\ne p$.
The momentum is bounded ($p \leq \kappa$) while energy remains unbounded, and the deformation does 
\textit{not} collapse to the SR form, in contrast to Eq.~\eqref{eq:ms_mdr}. The density of states behaves as 
$g(E)\,dE \propto p^2\,(dp/dE) = \kappa^2(1-e^{-E/\kappa})^2\,e^{-E/\kappa}\,dE$, 
which is exponentially suppressed at high energy. The partition 
function is therefore convergent without the need for a cutoff.

Three consequences for the present problem follow. First, the density of states is deformed at \textit{all} 
energies, $g(E)\,dE\propto p^2(E)\,(dp/dE)\,dE$ with $p(E)$ obtained from Eq.~\eqref{eq:ACdisp}, and the 
modification is not confined to a kinematic endpoint. Second, deviations from SR are controlled by the 
ratio $E/\kappa$ and therefore persist at temperatures well below the Planck scale, unlike the 
exponentially suppressed corrections found here. Third, because the density of states is exponentially 
suppressed at high energy rather than truncated at a kinematic endpoint of the MS type, the thermal 
integrals converge at all temperatures; the AC analogue of the equation of state nevertheless requires 
a separate calculation. A complete parallel thermodynamic 
treatment is thus not a by-product of the present analysis; we regard it as an important open problem and 
return to it in Sec.~\ref{sec:summary}.

\section{The Logarithmic Partition Function of DSR Photon Gas}
\label{sec:PF}

The statistical-mechanical construction used below rests on the following assumptions, which we state 
explicitly:
\begin{enumerate}
\item \textit{Additivity.} The total energy of a multiparticle state is the sum of the single-particle MS 
energies, $H=\sum_i E_i$, i.e.\ the deformed dispersion relation \eqref{eq:ms_mdr} is taken as the 
one-particle Hamiltonian and the deformed boost acts on it alone. 

\item \textit{Fock counting.} We employ standard bosonic Fock counting over 
independent modes together with the Bose--Einstein distribution at zero chemical potential, $\mu=0$. In the 
auxiliary (bicrossproduct) chart the spatial brackets of the deformed phase space are canonical and the 
time-translation generator $P_0$ satisfies the primitive coproduct
$\Delta(P_0) = P_0 \otimes \mathbf 1 + \mathbf 1 \otimes P_0$, so that distinct modes carry independent 
equilibrium occupation numbers and the grand-canonical construction is internally consistent; this does not 
demand that the nonlinear composition rule for physical observables close within a single mode. We emphasize 
that the primitive coproduct belongs to the auxiliary energy $e$, not to the MS energy $E$: since the map 
$e\mapsto E$ is nonlinear, $\Delta(E)$ is not primitive, and the additivity of assumption~1 is an explicit 
modelling assumption rather than a consequence of the coalgebra.
\end{enumerate}

Accordingly, the results obtained below should be read as the thermodynamics of the additive, hard-cutoff, 
effective model. What is specific to the MS model among the ingredients of the calculation is the covariant 
$E_{\rm P}$, i.e.\ the fact that the bound $E<E_{\rm P}$ is a kinematic statement valid in every inertial 
frame rather than a frame-dependent momentum-space cutoff.

For massless photons, Eq.~\eqref{eq:ms_mdr} reduces to the standard dispersion 
relation $E = pc$ (we retain $\hbar$, $c$, and $k_{\rm B}$ explicitly in order to display the 
standard thermodynamic results with their conventional dimensions in this section). Thus, 
within the MS model considered in this study, the photon dispersion relation itself remains 
unchanged. The DSR effect enters instead through the second invariant scale, which imposes 
an upper bound $E_{\rm P}$ on the single particle energy.

We now consider a photon gas confined to a three-dimensional volume $V$. For free particles, the 
number of states with momentum between $p$ and $p+dp$, including the two photon polarizations, is 
$\frac{V}{\pi^2 \hbar^3} p^2\, dp$. Using dispersion relation $E=pc$, the corresponding density of 
states in energy space reads $g(E) dE=\frac{V}{\pi^2 \hbar^3 c^3} E^2 dE$. According to the standard 
statistical mechanics (see, e.g., Ref.~\cite{KH1987}), and note the chemical potential for photon 
vanishes $\mu=0$, the grand partition function of a photon gas can be written as
\beq 
\label{eq:lnXi_sum}
 \ln \Xi = -\sum_E \ln \left(1-e^{-\beta E}\right) ~,
\eeq
where $\beta=1/(k_{\rm B}T)$. 
Replacing the sum in Eq. \eqref{eq:lnXi_sum} by an integral and imposing the DSR energy upper bound, we obtain
\beq 
\label{eq:lnXi_dsr1}
 \ln \Xi^{\rm (DSR)}= -\frac{V}{\pi^2 \hbar^3 c^3} \int_0^{E_{\rm P}} E^2 \ln\left(1-e^{-\beta E}\right) \, dE .
\eeq
Note that we have already imposed the upper bound $E_{\rm P}$ in the integral above. It is 
convenient to rewrite Eq.~\eqref{eq:lnXi_dsr1} by integration by parts:
\beq 
\label{eq:lnXi_parts}
\begin{aligned}
 \ln \Xi^{\rm (DSR)}=-\frac{V}{3\pi^2 \hbar^3 c^3}
 \Bigg[
 &\left. E^3 \ln\left(1-e^{-\beta E}\right)\right|_0^{E_{\rm P}}
 -\int_0^{E_{\rm P}} \frac{\beta E^3}{e^{\beta E}-1}\, dE
 \Bigg] ,
\end{aligned}
\eeq
where the boundary term at $E=0$ vanishes since $E^3\ln(1-e^{-\beta E}) \to 0$. Hence the
logarithmic partition function becomes
\beq 
\label{eq:lnXi_parts2}
\begin{aligned}
 \ln \Xi^{\rm (DSR)} = -\frac{V}{3 \pi^2 \hbar^3 c^3} \bigg[
 & E_{\rm{P}}^3 \ln \left(1-e^{-\beta E_{\rm{P}}}\right)
 +\int_{E_{\rm{P}}}^\infty \frac{\beta E^3}{e^{\beta E}-1}\, dE
 -\int_0^\infty \frac{\beta E^3}{e^{\beta E}-1} \, dE  \bigg] .
\end{aligned}
\eeq
Note that, to get Eq.~\eqref{eq:lnXi_parts2}, we rewrite the last terms in 
Eq.~\eqref{eq:lnXi_parts} as a subtraction of the last two terms in 
Eq.~\eqref{eq:lnXi_parts2}, so that the standard SR contribution (the last term in the above
equation) is clearly separated from the correction induced by the finite Planck-scale 
cutoff. In order to see this point, we introduce the dimensionless variables $x=\beta E$ and 
$x_{\rm P} \equiv \beta E_{\rm P}$, 
the logarithmic partition function can be written as
\beq 
\label{eq:lnXi_dsr_tail}
 \ln \Xi^{\rm (DSR)} =-\frac{V}{3\pi^2 \hbar^3 c^3 \beta^3}
    \left[x_{\rm{P}}^3 \ln \left(1-e^{-x_{\rm P}}\right)
    +\int_{x_{\rm P}}^\infty \frac{x^3}{e^x-1}\, dx -\zeta(4) \Gamma(4) \right] .
\eeq
In last equation, we have used the standard BE integral
\beq
\label{eq:zeta4}
 \Gamma(4)\zeta(4) = \int_0^\infty\frac{x^3}{e^x-1}\, dx = \frac{\pi^4}{15} ~,
\eeq
where $\zeta(s)$ is Riemann $\zeta$-function, defined as
\beq
 \zeta(s)=\frac{1}{\Gamma(s)} \int_0^\infty \frac{x^{s-1}}{e^x-1}\, dx ~,
\eeq
where $\Gamma(s)$ is the Gamma function, defined by the Euler integral
\beq
 \Gamma(n) = \int_0^\infty t^{n-1} e^{-t} \, dt , \qquad n > 0 ,
\eeq
which satisfies $\Gamma(n+1) = n!$ for integer $n$. In particular, for $n=4$, we have 
$\Gamma(4) = 3!$. One clearly identifies the standard SR photon gas contribution coming
from the last term in Eq.~\eqref{eq:lnXi_dsr_tail}.

The second term of Eq.~\eqref{eq:lnXi_dsr_tail} is incomplete BE integral (also referred 
to as the incomplete polylogarithm function, which is related to the standard 
polylogarithm~\cite{Abra1972}) with vanishing chemical potential:
\beq
\label{eq:Bxp}
 \mathcal{B} (y)\equiv \int_y^{\infty}\frac{x^3}{e^x-1}\, dx .
\eeq
However, for later convenience, and to simplify the subsequent calculation as well, 
in the following, we define an integral $\mathcal{J}$ as
\beq
\label{eq:Jxp}
 \mathcal{J} (y) \equiv \int_0^{y} \frac{x^3}{e^x-1}\, dx .
\eeq
Note that when $y \to \infty$, the limit of integral $\mathcal{J}$ is
\beq
\label{eq:Jinf}
 \mathcal{J} (y \to \infty) = \Gamma(4)\zeta(4) = \frac{\pi^4}{15} ~.
\eeq
The integral $\mathcal{J}$ and its property, Eq.~\eqref{eq:Jinf}, will be frequently used 
in the evaluation of the thermodynamic quantities.

With the definition of integral $\mathcal{J}$ in Eq.~\eqref{eq:Jxp}, now
Eq.~\eqref{eq:lnXi_dsr_tail} becomes
\beq 
\label{eq:lnXi_dsr}
 \ln \Xi^{\rm (DSR)} = \frac{V}{3\pi^2 \hbar^3 c^3 \beta^3}
 \left[ \mathcal{J} (x_{\rm P})-x_{\rm{P}}^3\ln\left(1-e^{-x_{\rm P}}\right) \right] .
\eeq
In the limit $x_{\rm{P}} \to \infty$, the boundary term 
$x_{\rm{P}}^3\ln(1-e^{-x_{\rm{P}}})$ vanishes, and applying Eq.~\eqref{eq:Jinf}, 
Eq.~\eqref{eq:lnXi_dsr} therefore reduces to
\beq 
\label{eq:lnXi_sr}
 \ln \Xi^{\rm (SR)}=\frac{\pi^2V}{45(\beta\hbar c)^3} = \frac{\pi^2V}{45(\hbar c)^3} (k_{\rm B} T)^3 ,
\eeq
which is exactly the standard SR logarithmic partition function for a photon gas.

\section{Thermodynamic Quantities of DSR Photon Gas}
\label{sec:thermo}

Having derived the explicit expression for the grand partition function of a photon gas in 
the preceding section, we now proceed to compute the corresponding thermodynamic quantities. 
From standard statistical mechanics, the Helmholtz free energy $F$, internal energy $U$, 
heat capacity $C_{\rm V}$, entropy $S$, and pressure $P$ are given by the following derivatives 
of the logarithm of the grand partition function:
\begin{align}
  F^{\rm (DSR)} &= -\frac{1}{\beta} \ln \Xi^{\rm (DSR)} , \label{eq:Fdsr} \\
  U^{\rm (DSR)} &= -\frac{\partial}{\partial \beta} \ln \Xi^{\rm (DSR)} , \label{eq:Udsr} \\
  C^{\rm (DSR)}_V &= \left(\frac{\partial U^{\rm (DSR)}}{\partial T}\right)_{\rm V} , \label{eq:Cdsr} \\
  S^{\rm (DSR)} &= -\left(\frac{\partial F^{\rm (DSR)}}{\partial T}\right)_{\rm V} , \label{eq:Sdsr} \\
  P^{\rm (DSR)} &= -\left(\frac{\partial F^{\rm (DSR)}}{\partial V}\right)_{\rm T} . \label{eq:Pdsr} 
\end{align}

In the following derivation, we retain $\hbar$, $c$, and $k_{\rm B}$ explicitly where needed, 
so that the recovery of the standard SR results can be clearly exhibited and directly compared 
with the DSR expressions.

\subsection{Helmholtz Free Energy of DSR Photon Gas}

From Eqs.~\eqref{eq:lnXi_dsr} and \eqref{eq:Fdsr}, the Helmholtz free energy is obtained as follows:
\beq
\label{eq:F_dsr}
 F^{\rm (DSR)}  =\frac{V}{3\pi^2 \hbar^3 c^3 \beta^{4}} \left[ x_{\rm P}^3 \ln\left(1-e^{-x_{\rm P}}\right)
 -\mathcal{J} (x_{\rm P}) \right] ~,
\eeq
where $x_{\rm P} \equiv \beta E_{\rm{P}}$ and $\mathcal{J} (x_{\rm P})$ is the incomplete BE integral defined 
in Eq.~\eqref{eq:Jxp}. In the limit $x_{\rm{P}} \to \infty$, the first boundary term 
$x_{\rm{P}}^3\ln(1-e^{-x_{\rm P}})$ vanishes, and the second term, according to Eq.~\eqref{eq:Jinf}, 
$\mathcal{J} (x_{\rm P} \to \infty) = \pi^4/15$, such that Eq.~\eqref{eq:F_dsr} reduces to
\beq
 F^{\rm (SR)} = -\frac{\pi^2 V}{45\hbar^3 c^3}(k_{\rm B} T)^4 ,
\eeq
which recovers the standard SR expression for free energy of a photon gas.

The recovery of the SR photon gas expression also highlights the discrepancy with the earlier 
calculation in Ref.~\cite{Das2010}. The free energy and other thermodynamic 
quantities obtained in their work do not reduce to the standard SR expressions in the corresponding 
limit, instead, their results yield the classical ideal gas form. This discrepancy arises because 
they employed the MB distribution, which is applicable only 
to classical ideal gases, rather than the BE statistics required for photons.

\begin{figure}[h!]
\centering
\includegraphics[width=0.72\linewidth]{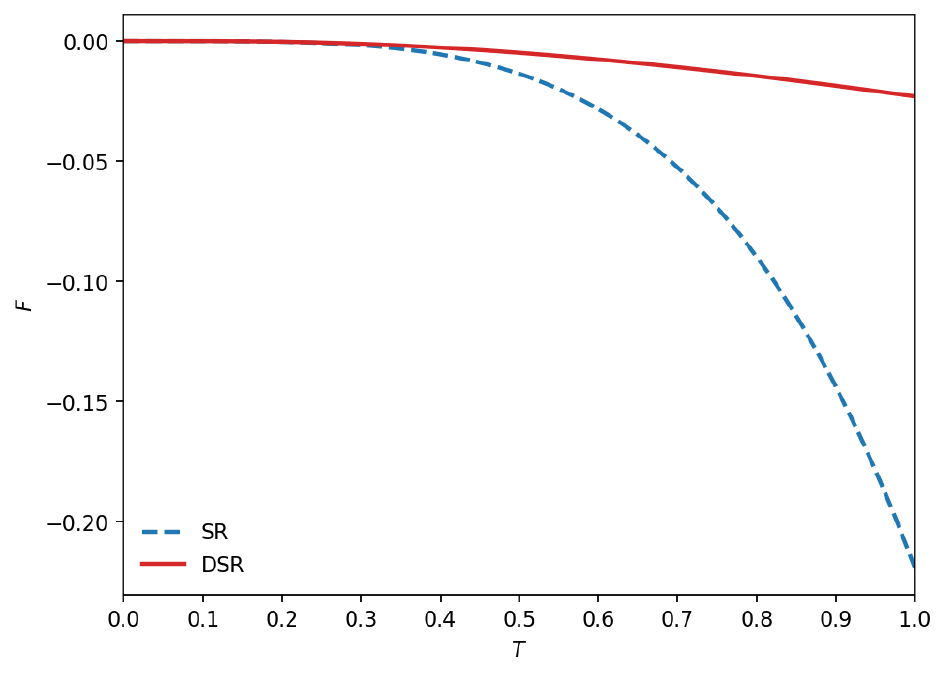}
\caption{(Color online) The Helmholtz free energy $F$ is plotted as a function of the temperature 
$T$ for the SR (blue dashed line), DSR (red solid line), respectively (in units where 
$E_{\rm P} = c = \hbar = k_{\rm B} = V = 1$).}
\label{fig:f-t}
\end{figure}

Fig.~\ref{fig:f-t} shows the temperature dependence of the Helmholtz free energy in both the 
SR and the DSR MS models. In the numerical evaluation we adopt Planck units, setting 
$E_{\rm{P}} = c = \hbar= k_{\rm B}=1$ and unit volume, so that $T=1$ corresponds to the Planck 
temperature $T_{\rm P}$. 
At low temperatures (relative to $T_{\rm P}$), the thermal population of high-energy modes is 
exponentially suppressed by the Boltzmann factor, so states near the cutoff contribute negligibly 
and the two curves are almost indistinguishable. The DSR curve begins to depart visibly from the 
SR result only when the temperature becomes a non-negligible fraction of the Planck temperature. 
As $T$ approaches $T_{\rm P}$, the finite upper energy bound removes an increasingly significant 
portion of the high-energy tail, and the magnitude of the free energy is therefore substantially 
reduced.

\subsection{Internal Energy of DSR Photon Gas}

Another important thermodynamic quantity, the internal energy $U^{\rm (DSR)}$ of DSR photon gas, 
can also be derived from 
the partition function. Combining Eqs.~\eqref{eq:lnXi_dsr} and \eqref{eq:Udsr}, we obtain 
the internal energy $U^{\rm (DSR)}$ of DSR photon gas as
\begin{eqnarray}
 U^{\rm (DSR)} &=& -\frac{V}{3\pi^2 \hbar^3 c^3\beta^4}\left[-3\mathcal{J} (x_{\rm P})
 +\frac{x_{\rm{P}}^4}{e^{x_{\rm P}}-1}-\frac{x_{\rm P}^4}{e^{x_{\rm P}}-1} \right] \nonumber \\
 &=& \frac{V}{\pi^2 \hbar^3 c^3 \beta^4} \, \mathcal{J} (x_{\rm P}) , \label{eq:U_dsr}
\end{eqnarray}
where we have applied the derivative relations:
\beq
\label{eq:dBdx}
 \frac{d\mathcal{J} (x_{\rm P})}{dx_{\rm P}}= \frac{x_{\rm P}^3}{e^{x_{\rm P}}-1} , 
 \qquad \frac{dx_{\rm P}}{d\beta}= \frac{x_{\rm{P}}}{\beta} .
\eeq                         
Also, note that the final compact expression for the internal energy $U^{\rm (DSR)}$ follows 
because the last two terms in the bracket: one arising from the upper cutoff and the other 
from the boundary term, cancel each other exactly.

Similar to the case of free energy, in the limit $x_{\rm P} \to \infty$, with Eq.~\eqref{eq:Jinf},
we have Eq.~\eqref{eq:U_dsr} reduced to
\beq
 U^{\rm (SR)} =\frac{V}{\pi^2 \hbar^3 c^3\beta^4}\frac{\pi^4}{15}
 =\frac{\pi^2 V}{15\hbar^3c^3} (k_{\rm B} T)^4 ~,
\eeq
which recovers the standard SR internal energy for a photon gas, as expected.

\begin{figure}[h!]
\centering
\includegraphics[width=0.72\linewidth]{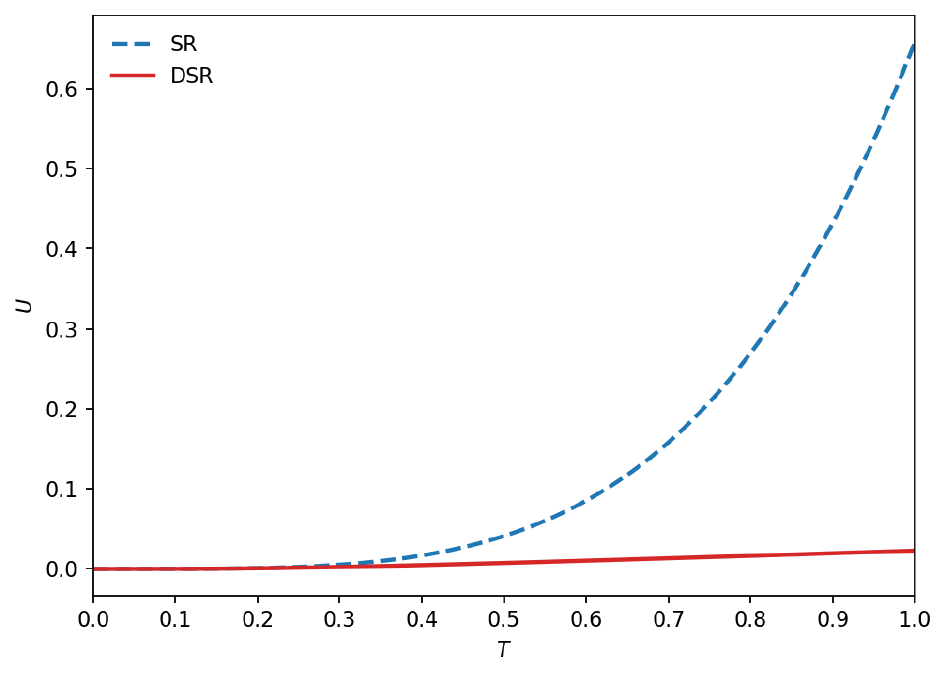}
\caption{(Color online) Temperature dependence of the internal energy $U$ of a photon gas, plotted 
in the same manner as in Fig.~\ref{fig:f-t}. The DSR result $U^{\rm (DSR)}$ (red solid line) is 
significantly suppressed relative to the standard SR result $U^{\rm (SR)}$ (blue dashed line), due 
to the exclusion of high-energy photon modes above the cutoff $E_{\rm{P}}$ (in units where 
$E_{\rm P} = c = \hbar = k_{\rm B} = V = 1$).}
\label{fig:u-t}
\end{figure}

The corresponding numerical results are displayed in Fig.~\ref{fig:u-t}, which shows the temperature 
dependence of the internal energy in both the SR and the DSR MS models. It is evident from the figure 
that the DSR internal energy grows more slowly with temperature than its SR counterpart. The two 
curves begin to deviate noticeably at around $T \simeq 0.2\,T_{\rm P}$, and the deviation becomes 
increasingly pronounced as the temperature rises. This behavior originates from the finite upper bound 
$E_{\rm P}$ imposed on single-particle energies in the DSR framework, which reduces the number of 
microscopic states accessible to the photon gas compared with the SR case. Consequently, the internal 
energy in the MS model is systematically smaller than that in the SR case.

It is also worth noting that, in standard SR photon gas thermodynamics, the Helmholtz free energy and 
the internal energy satisfy the well-known relation
\beq
 F^{\rm (SR)} = -\frac{1}{3} U^{\rm (SR)} ,
\eeq
which is a characteristic feature of photon gas thermodynamics. In the MS model of DSR, however, this 
relation no longer holds, as can be directly verified from the expressions derived in 
Eqs.~\eqref{eq:F_dsr} and \eqref{eq:U_dsr}. The breakdown of this relation arises because the finite 
cutoff introduces an additional boundary term, 
$\frac{V x_{\rm P}^3}{3\pi^2 \hbar^3 c^3 \beta^{4}} \ln \left(1 - e^{-x_{\rm P}}\right)$ 
in the free energy, whereas the internal energy is determined solely by the integral 
$\mathcal{J}(x_{\rm P})$. This extra term in the free energy will further contribute to the pressure 
and modify the pressure to energy-density relation, as we will show in Sec.~\ref{sec:eos}.

\subsection{Entropy of DSR Photon Gas}

After obtaining the Helmholtz free energy $F^{\rm(DSR)}$, the entropy of the photon gas follows 
directly from Eq.~\eqref{eq:Sdsr}. For convenience, we change the derivative with respect to 
$\beta$, instead of $T$, i.e., $S^{\rm (DSR)}= -\left(\frac{\partial F^{\rm (DSR)}}{\partial T}\right)_V
= k_{\rm B}\beta^2 \left(\frac{\partial F^{\rm (DSR)}}{\partial\beta}\right)_V$. As a result, 
the entropy is
\beq
\begin{aligned}
 S^{\rm (DSR)} &= k_{\rm B}\beta^2 \frac{\partial}{\partial \beta} \left\{
 \frac{V}{3\pi^2 \hbar^3 c^3} \left[ \beta^{-4} x_{\rm{P}}^3 \ln\left(1-e^{-x_{\rm P}}\right)
 -\beta^{-4}\mathcal{J} (x_{\rm P}) \right] \right\} \\
 &= \frac{Vk_{\rm B}\beta^2}{3 \pi^2 \hbar^3 c^3\beta^{5}} \left[-x_{\rm{P}}^3 \ln\left(1-e^{-x_{\rm P}}\right)
    +\frac{x_{\rm{P}}^4}{\left(e^{x_{\rm P}}-1 \right)} + 4\mathcal{J} (x_{\rm P})
    -\frac{x_{\rm{P}}^4}{\left(e^{x_{\rm P}}-1 \right)} \right] \\
 &= \frac{Vk_{\rm B}}{3\pi^2 \hbar^3 c^3 \beta^3} \left[ 4\mathcal{J} (x_{\rm P})
    -x_{\rm{P}}^3\ln\left(1-e^{-x_{\rm P}}\right) \right] , \label{eq:S_dsr}
\end{aligned}
\eeq
where we have applied the derivative relations in Eq.~\eqref{eq:dBdx}.

Similarly, in the limit $E_{\rm{P}}\to \infty$ at finite temperature, applying Eq.~\eqref{eq:Jinf},
and $x_{\rm{P}}^3 \ln \left(1-e^{-x_{\rm P}} \right) \to 0$, such that the entropy reduces 
to the standard special-relativistic result,
\beq
 S^{\rm (SR)} = \frac{4\pi^2 V}{45\hbar^3c^3} k_{\rm B}^4 T^3 ,
\eeq
which is exactly the expression in the special-relativistic limit.

\begin{figure}[h!]
\centering
\includegraphics[width=0.72\linewidth]{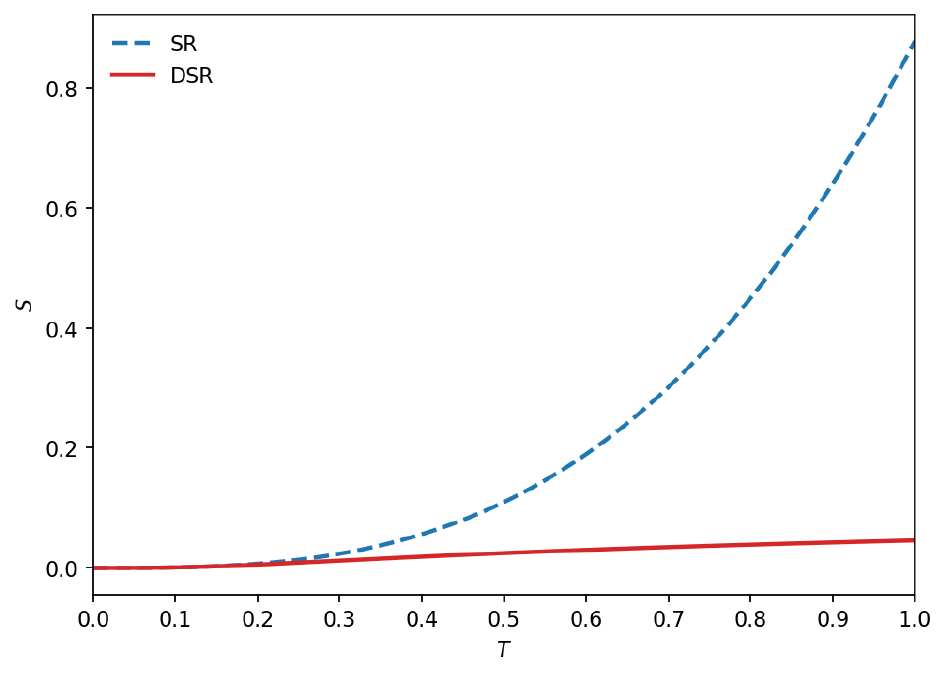}
\caption{(Color online) Similar to Fig.~\ref{fig:f-t}, but for the entropy $S$ of the photon gas. 
The DSR entropy (red solid line) is again significantly smaller than the SR entropy (blue dashed line) 
at high temperature because the finite cutoff reduces the available phase space (in units where 
$E_{\rm P} = c = \hbar = k_{\rm B} = V = 1$).}
\label{fig:s-t}
\end{figure}

Fig.~\ref{fig:s-t} shows the temperature dependence of the entropy of the photon gas in SR 
and in the DSR MS model. Similar to the cases of the Helmholtz free energy and the internal
energy, the entropy in the MS model increases more slowly with temperature than its SR counterpart. 
Since entropy provides a direct measure of the logarithm of the number of accessible microscopic 
states, this result indicates that the presence of the upper energy bound $E_{\rm{P}}$ in the MS 
model reduces the number of photon states available to the system relative to the SR case. As the 
temperature rises, high-energy photon modes become increasingly populated in SR, whereas in the 
MS model all states with $E> E_{\rm P}$ are strictly excluded. As a result, the deviation between 
the two curves becomes increasingly pronounced at high temperatures, consistent with the trends 
discussed above for the free energy and internal energy. 

It is interesting to compare this result with that obtained in the Lorentz-violating model of 
Ref.~\cite{Cama2007}. 
In that case, the MDR leads to an entropy \textit{larger} than the SR result, 
implying that the number of accessible photon microstates is enhanced relative to the 
SR case. By contrast, in the MS model considered here, the finite 
upper energy bound suppresses the entropy. Nevertheless, in both cases the entropy vanishes as 
$T\to 0$, in accordance with the third law of thermodynamics.

\subsection{Equation of State of DSR Photon Gas}
\label{sec:eos}

The pressure can be obtained from the Helmholtz free energy through 
Eq.~\eqref{eq:Pdsr}. Since $F^{\rm (DSR)}$ is proportional to the volume $V$, and 
$x_{\rm P}=\beta E_{\rm{P}}$ is independent of $V$, the pressure is simply 
\beq 
\label{eq:P_dsr}
 P^{\rm (DSR)} = \frac{1}{3\pi^2 \hbar^3 c^3 \beta^4} \left[\mathcal{J} (x_{\rm P}) 
 - x_{\rm{P}}^3 \ln \left(1-e^{-x_{\rm P}}\right) \right] .
\eeq

Again, in the limit $E_{\rm{P}} \to \infty$, Eq.~\eqref{eq:P_dsr} reduces to
\beq
 P^{\rm (SR)} = \frac{1}{3\pi^2 \hbar^3 c^3 \beta^4}\frac{\pi^4}{15} = \frac{\pi^2}{45\hbar^3c^3}
 (k_{\rm B} T)^4 ~,
\eeq
which is exactly the expression recovered in the special-relativistic limit.

For an ordinary photon gas, it is well-known that the equation of state satisfies relation $P=U/(3V)$. 
In the present MS model, however, we find
\beq
 P^{\rm (DSR)}
 =\frac{U^{\rm (DSR)}}{3V}-\frac{k_{\rm B} T E_{\rm P}^3}{3 \pi^2 \hbar^3 c^3} 
 \ln \left(1-e^{-E_{\rm P}/k_{\rm B} T}\right) ,
\label{eq:eos_dsr}
\eeq
where we have explicitly expressed in terms of $E_{\rm P}$ and $T$.  
That is to say, the standard photon gas equation of state receives a modification in the form of a 
boundary term arising from the finite upper bound $E_{\rm P}$. 
Since $\ln(1-e^{-x_{\rm P}})<0$, this additional term is positive. Nevertheless, for finite $E_{\rm{P}}$, 
$P^{\rm (DSR)}$ remains smaller than 
$P^{\rm (SR)}$ in its value, because the restricted phase space reduces the total number of 
contributing modes compared with the unbounded SR case.

\begin{figure}[h!]
\centering
\includegraphics[width=0.72\linewidth]{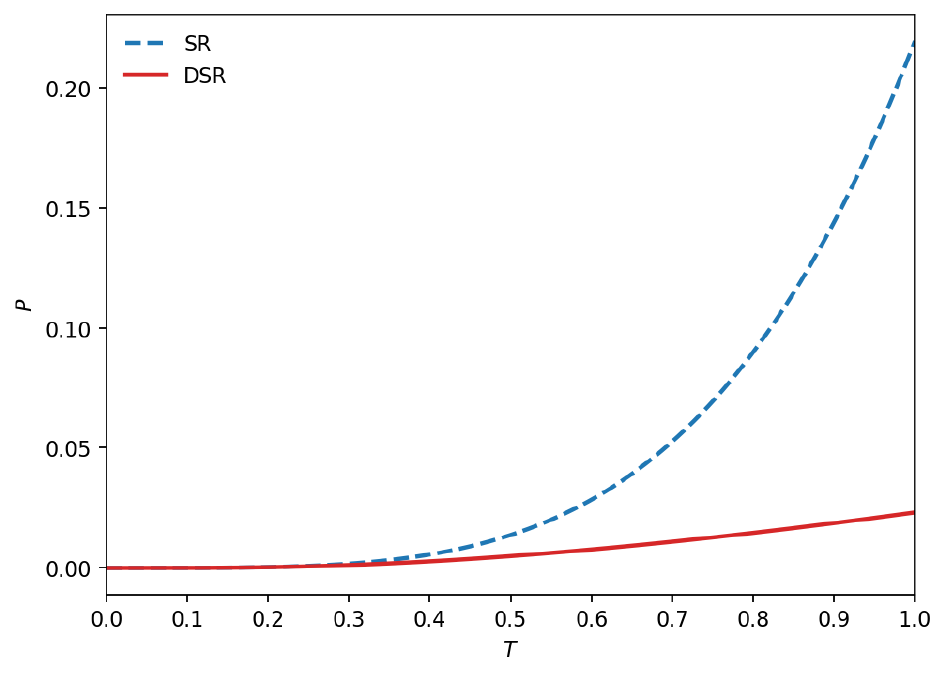}
\caption{(Color online) Similar to the Fig.~\ref{fig:f-t}, but for the pressure $P$ of the photon gas. 
The finite energy cutoff greatly lowers the DSR pressure (red solid line) relative to the standard SR 
result (blue dashed line) near the Planck temperature (in units where 
$E_{\rm P} = c = \hbar = k_{\rm B} = V = 1$).}
\label{fig:p-t}
\end{figure}

We plot the pressure $P$ as a function of temperature $T$ in Fig.~\ref{fig:p-t}. 
As in the case of the internal energy, the DSR pressure follows the SR curve at low temperatures and 
becomes increasingly suppressed as the finite upper energy bound becomes thermally relevant. 
This behavior is consistent with the interpretation that the Planck-scale upper bound reduces the 
number of photon modes available to contribute to the radiation pressure. Such a modification is 
significant only in extreme high temperature environments. For ordinary temperatures, the correction 
remains far beyond observable reach.

It is useful to isolate the modification of the equation of state by introducing the internal energy
density $u\equiv U/V$. From Eqs.~\eqref{eq:U_dsr} and \eqref{eq:P_dsr}, one obtains
\beq 
\label{eq:P-u}
 \left( \frac{P}{u}\right)^{\rm (DSR)} =
 \frac{1}{3}\left[ 1-\frac{x_{\rm P}^3\ln\left(1-e^{-x_{\rm P}}\right)}{\mathcal{J} (x_{\rm P})} \right],
\eeq
or, explicitly express the equation of state in terms of $T$ and $E_{\rm P}$ as 
\beq 
\label{eq:P-u-exp}
 \left( \frac{P}{u}\right)^{\rm (DSR)} =
 \frac{1}{3}\left[ 1-\left(\frac{E_{\rm P}}{k_{\rm B} T}\right)^3 
 \frac{\ln \left(1-e^{-E_{\rm P}/k_{\rm B}T}\right)}{\mathcal{J} \left(E_{\rm P}/k_{\rm B} T \right)} \right] .
\eeq
The second term in Eq.~\eqref{eq:P-u} isolates the extra term contributing to the pressure to energy 
density ratio. This additional contribution modifies the equation of state of the photon gas 
relative to the standard SR form and may potentially manifest itself in other extreme astrophysical 
environments, thereby warranting further investigation.

On the other hand, the SR value $P/u=1/3$ is recovered in the low temperature limit. Indeed, when $x_{\rm P}\gg 1$,
$\mathcal{J} (x_{\rm P}) \to \pi^4/15$ and $\ln(1-e^{-x_{\rm P}})\simeq -e^{-x_{\rm P}}$, so that
\beq 
\label{eq:Pulim}
 \frac{P^{\rm (DSR)}}{u^{\rm (DSR)}}-\frac{1}{3} \simeq \frac{x_{\rm P}^3e^{-x_{\rm P}}}{3\pi^4/15} \to 0 ~.
\eeq

\begin{figure}[h!]
\centering
\includegraphics[width=0.72\linewidth]{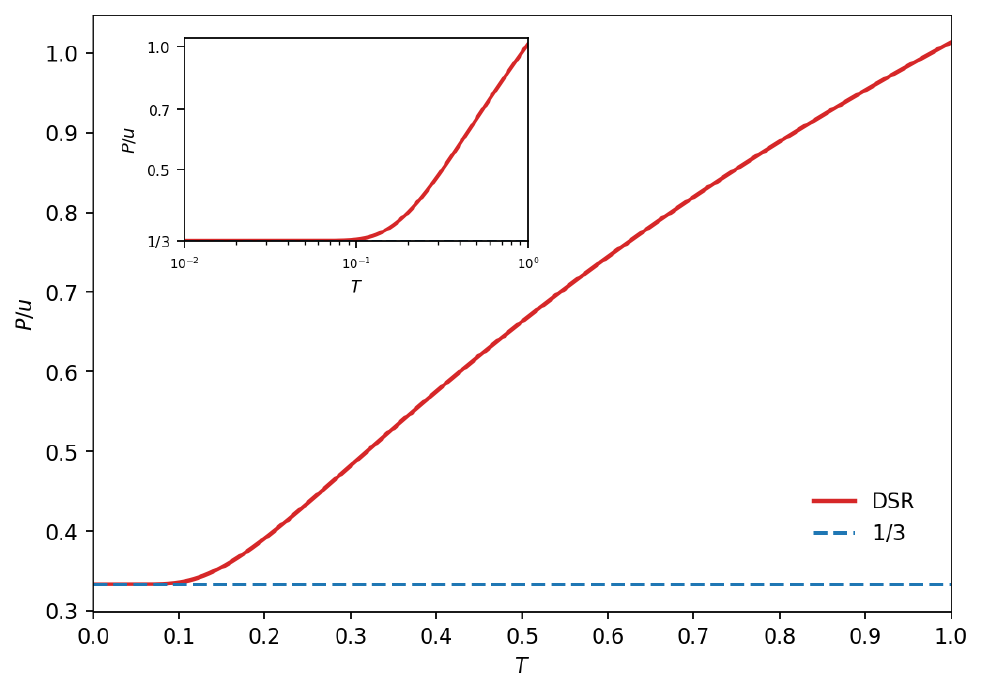}
\caption{Equation of state ratio $P/u$ as a function of temperature in the MS model of DSR. 
The dashed horizontal line denotes the standard SR value $P/u=1/3$. The main panel shows the 
behavior up to the Planck temperature, while the inset displays the low-temperature region on 
a semi-log plot. At low temperatures the DSR correction is exponentially suppressed, and the 
ratio is nearly indistinguishable from $1/3$. A visible upward deviation appears only when 
the temperature becomes a sizable fraction of the Planck temperature $T_{\rm P}$ (in units where 
$E_{\rm P} = c = \hbar = k_{\rm B} = V = 1$).}
\label{fig:p-u-t}
\end{figure}

Fig.~\ref{fig:p-u-t} shows the temperature dependence of the pressure to energy density ratio $P/u$ 
of DSR photon gas. 
In the main panel, we show the $P/u$ ratio from $T=0$ to $T=1$. In order to see clearly $P/u$ behavior 
in low temperature regime, an inset plot zooms in on the low 
temperature region by using a logarithmic temperature axis.
At very low temperatures, the DSR curve is essentially 
indistinguishable from the SR value $1/3$, as expected from the exponentially suppressed correction 
in Eq.~\eqref{eq:Pulim}. A visible upward deviation appears only when the temperature 
becomes a non-negligible fraction of the Planck temperature, around $T/T_{\rm P} \sim 10^{-1}$ 
(see the inset plot of Fig.~\ref{fig:p-u-t}). 
As $T$ approaches $T_{\rm P}$, the cutoff boundary term becomes increasingly important, and the 
ratio $P/u$ departs significantly from the standard radiation value.

The modification of the equation of state found above deserves a careful discussion of its origin and of 
the domain in which our description may be trusted. It is convenient to take the formal limit 
$x_{\rm P}\to 0$ of Eq.~\eqref{eq:P-u-exp}, for which $\mathcal{J}(x_{\rm P})\simeq x_{\rm P}^3/3$ and 
$\ln(1-e^{-x_{\rm P}})\simeq\ln x_{\rm P}$. One then obtains
\beq
\label{eq:Pudiv}
 \frac{P}{u}\simeq\frac{1}{3}\left(1-3\ln x_{\rm P}\right) \;\xrightarrow[\;x_{\rm P}\to0\;]{}\;+\infty ,
\eeq
so that the ratio diverges logarithmically: the qualitative behaviour is not a mild departure from 
the radiation value but an unbounded growth. The energy density, by contrast, grows only linearly,
\beq
\label{eq:ugrowth}
 u = \frac{U}{V} \simeq \frac{E_{\rm P}^3 k_{\rm B}T}{3\pi^2\hbar^3c^3} \qquad (k_{\rm B}T\gg E_{\rm P}) ,
\eeq
which is the statement, in terms of $\mathcal{J}(x_{\rm P})\simeq x_{\rm P}^3/3$, that the heat capacity 
saturates to the constant of Eq.~\eqref{eq:Cvsat}. 
In the formal high-temperature limit, $u$ grows linearly with $T$, whereas $P$ grows as $T \ln(T/T_{\rm P})$. 
Consequently, both $P/u$ and $dP/du$ grow logarithmically. 
The sound speed is ($c=1$)
\beq
\label{eq:cs}
 c_s^2 = \left(\frac{\partial P}{\partial u}\right)_{\rm S} \simeq \frac{1}{3}\left(4-3\ln x_{\rm P}\right)
\;\xrightarrow[\;x_{\rm P} \to 0\;]{}\;+\infty
\eeq
diverges logarithmically as $x_{\rm P}\to 0$. If the standard relativistic causality bound $c_s\le c$ is 
adopted, this bound is formally violated. The trace of the stress–energy tensor, $u-3P\simeq 3u\ln x_{\rm P}$, 
also becomes arbitrarily negative in this limit.

However, several points must be emphasized. First, these features are \textit{not} predictions of DSR. 
None of Eqs.~\eqref{eq:Pudiv}--\eqref{eq:cs} depends on the deformed boost; they follow solely from truncating 
the energy integral at a finite upper limit while retaining the flat counting measure. Note also that, under 
the invariant measure, the situation is still worse, since the partition function does not exist at all. 
We therefore do not claim $P/u > 1/3$ at $T \gtrsim E_{\rm P}$ as a signature of MS kinematics. Second, they 
indicate that the pressure prescription is incomplete. The origin is that the truncation destroys the conformal 
invariance of the free photon gas: the boundary term $-x_{\rm P}^3\ln(1-e^{-x_{\rm P}})$ is a surface 
contribution to $F$ that has no counterpart in $U$ --- which is why $F\neq -U/3$ --- and it is this surface 
term that diverges. A prescription in which the pressure is defined thermodynamically, 
$P=-(\partial F/\partial V)_T$, while the spectrum is truncated at a fixed energy. 
The fixed upper energy bound introduces a boundary contribution in the thermodynamic pressure. Relating this 
contribution to mechanical pressure requires specifying how the mode spectrum and the cutoff are implemented 
when the volume changes.
Third, as an \textit{effective} model, the range of validity is confined to $T \lesssim E_{\rm P}$. We therefore 
regard the $T\gtrsim E_{\rm P}$ region as lying outside the domain of validity of the description rather than as a 
prediction, and we attach no physical interpretation to it.

\subsection{Heat Capacity of DSR Photon Gas}
\label{sec:CV}

The heat capacity at constant volume can be obtained from the internal energy via Eq.~\eqref{eq:Cdsr}, 
or for convenience, via $C_V =-k_{\rm B}\beta^2 \left(\frac{\partial U}{\partial\beta}\right)_V$. 
Thus we obtain
\beq
\label{eq:Cv_dsr}
 C_{\rm V}^{\rm (DSR)} = -k_{\rm B}\beta^2 \frac{\partial}{\partial\beta} \left[
 \frac{V}{\pi^2 \hbar^3 c^3 \beta^4} \mathcal{J} (x_{\rm P}) \right] 
 = \frac{V k_{\rm B}}{\pi^2 \hbar^3 c^3 \beta^3} \left[4\mathcal{J} (x_{\rm P})
 - \frac{x_{\rm P}^4}{e^{x_{\rm P}}-1} \right] ,
\eeq
where $\mathcal{J} (x_{\rm P})$ is the integral defined in Eq.~\eqref{eq:Jxp}. In deriving 
Eq.~\eqref{eq:Cv_dsr}, we used the Eq.~\eqref{eq:dBdx}. 
In the limit $E_{\rm{P}} \to \infty$, Eq.~\eqref{eq:Cv_dsr} reduces to
\beq
 C_{\rm V}^{\rm (SR)} = \frac{4\pi^2 V}{15\hbar^3 c^3} k_{\rm B}^4 T^3 ,
\eeq
which is exactly the standard SR heat capacity, as expected.

\begin{figure}[h!]
\centering
\includegraphics[width=0.72\linewidth]{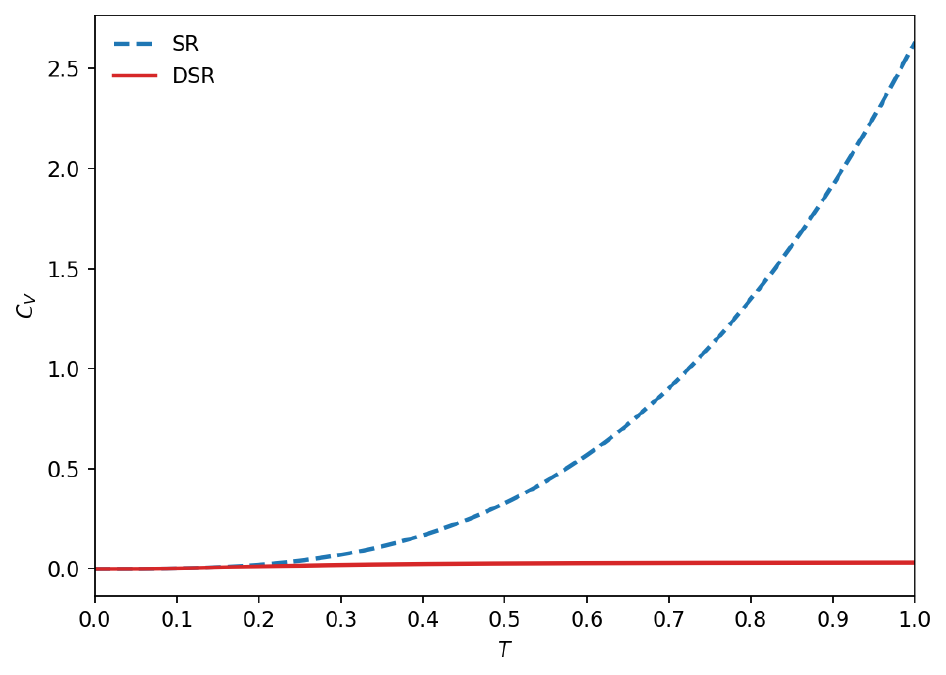}
\caption{Heat capacity $C_{\rm V}$ of the photon gas as a function of temperature $T$. The DSR curve 
(red solid line) is significantly suppressed relative to the SR behavior (blue dashed line), as the 
finite upper energy bound becomes thermally relevant (in units where 
$E_{\rm P} = c = \hbar = k_{\rm B} = V = 1$).}
\label{fig:cv-t}
\end{figure}

In Fig.~\ref{fig:cv-t}, we show the temperature dependence of the heat capacity $C_{\rm V}$ 
of the photon gas in the SR (blue dashed line) and in the DSR MS model (red solid line). 
As with the other thermodynamic quantities discussed above, the heat capacity in the MS 
model increases more slowly with temperature than in the SR case. At low temperatures, the 
two curves are almost indistinguishable, since the thermal population of modes near the 
cutoff is exponentially suppressed. As the temperature increases toward the Planck scale, 
the finite upper energy bound becomes increasingly important, and the DSR curve deviates 
significantly from the SR result.

It is also instructive to examine the formal high temperature behavior of 
Eq.~\eqref{eq:Cv_dsr}. This limit corresponds to $x_{\rm P} = \beta E_{\rm P} \ll 1$, 
or equivalently $k_{\rm B} T \gg E_{\rm P}$. Although this regime lies beyond the 
temperature range shown in Fig.~\ref{fig:cv-t}, it reveals an important consequence of 
the finite cutoff. For $x_{\rm P} \ll 1$, we have $\mathcal{J}(x_{\rm P}) \simeq x_{\rm P}^3/3$ 
and $\frac{x_{\rm P}^4}{e^{x_{\rm P}} - 1} \simeq x_{\rm P}^3$. Substituting these into 
Eq.~\eqref{eq:Cv_dsr} yields
\beq
\label{eq:Cvsat}
 C_{\rm V}^{\rm (DSR)} \to \frac{V k_{\rm B}}{3\pi^2 \hbar^3 c^3} E_{\rm P}^3 , \qquad x_{\rm P} \to 0 ~.
\eeq
Thus, Eq.~\eqref{eq:Cvsat} gives the formal $x_P\to 0$ limit of the hard-cutoff model. In contrast to 
the SR heat capacity, which grows as $\sim T^3$, the DSR heat capacity approaches a finite constant in 
the regime $k_{\rm B} T \gg E_{\rm P}$ (however, the physical applicability in this regime remains 
unestablished). 
This saturation effect reflects the finite phase space volume available below the upper bound $E_{\rm P}$.

We emphasise that all apparent effects for \(T\gtrsim E_{\rm P}\), including the suppression pattern, 
heat-capacity saturation, and modified equation of state, are artefacts introduced by the hard cutoff. 
They belong to the hard-cutoff prescription and not to MS kinematics.

\begin{figure}[h!]
\centering
\includegraphics[width=0.72\linewidth]{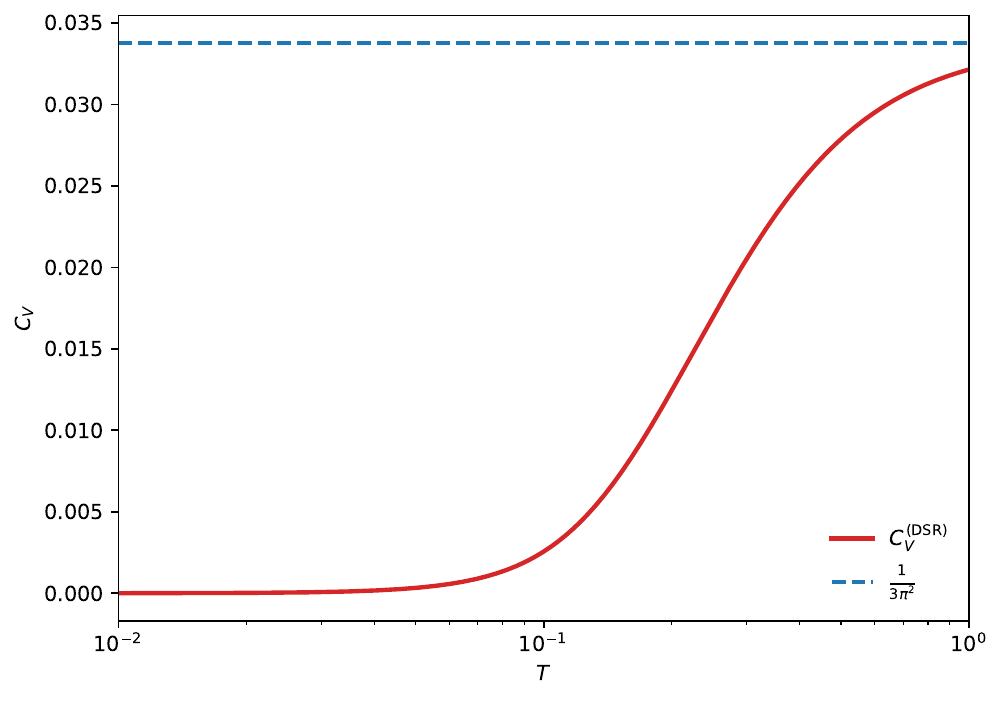}
\caption{The DSR heat capacity $C_{\rm V}^{(\rm DSR)}$ of the photon gas (red solid line) is shown 
as a function of temperature $T$. It tends to zero at the low-temperature end of the displayed range, 
vanishing as $T \to 0$ in accordance with the third law. In the high-temperature limit $T \to \infty$, 
the heat capacity approaches a saturated value given by $\frac{1}{3\pi^2}$ (in units where 
$E_{\rm P} = c = \hbar = k_{\rm B} = V = 1$).}
\label{fig:cv-t2}
\end{figure}

\section{Summary and Conclusions}
\label{sec:summary}

In this work, we have investigated the thermodynamic properties of a photon gas within the 
framework of doubly special relativity, focusing on the Magueijo-Smolin (MS) model. 
In this realization, although the dispersion relation is generally modified to
$E^2 - p^2 = m^2 \left(1 - \frac{E}{E_{\rm P}}\right)^2$,
photons obey the standard dispersion relation $E = pc$ on account of their vanishing mass. 
The Planck scale effect thus enters through the observer-independent upper bound $E_{\rm P}$
on photon energy. Consequently, the modification in the statistical calculation is 
implemented by replacing the upper limit of the energy integral with $E_{\rm P}$. 
It should be emphasized that DSR itself does not uniquely determine the correct modified 
dispersion relation; different DSR models can lead to different deformations, and 
the physically realized form, if any, must ultimately be constrained by experiments.

Three logically distinct ingredients enter the calculation, and it is important to keep them apart. 
The first is the Bose-Einstein statistics with vanishing chemical potential ($\mu=0$), which is 
required for photons and whose omission in earlier treatments prevented the recovery of the 
special-relativistic limit. The second is the truncation of the single-particle spectrum at 
$E_{\rm P}$, which is responsible for the suppression of the thermodynamic functions and which 
would arise for any hard upper bound on the photon energy. The third is the covariant status of 
$E_{\rm P}$: it is a second invariant of the deformed boost rather than an ad hoc regulator, so 
that the bound $E<E_{\rm P}$ holds for all inertial observers. 

Adopting Bose-Einstein statistics with $\mu=0$ and the flat counting phase-space measure, we derived 
the grand partition function of the photon gas and obtained the corresponding thermodynamic quantities, 
including the Helmholtz free energy, internal energy, entropy, pressure, and heat capacity. We showed 
explicitly that all these expressions reduce to the standard special-relativistic photon gas formulae 
in the limit $E_{\rm P} \to \infty$. 
At temperatures approaching the Planck scale, however, the finite upper energy bound leads to 
significant deviations from the special-relativistic results: the absolute value of Helmholtz free 
energy, internal energy, entropy, pressure, and heat capacity are all substantially suppressed relative 
to their SR counterparts, and the equation of state, i.e., the usual pressure-density relation $P/u = 1/3$, 
is modified by a cutoff-dependent term
$-\left(E_{\rm P}/k_{\rm B} T\right)^3 
\ln \left(1-e^{-E_{\rm P}/k_{\rm B}T}\right)/3\mathcal{J} \left(E_{\rm P}/k_{\rm B}T \right)$.
The numerical results demonstrate these features explicitly: at low temperatures the DSR and SR curves 
are nearly indistinguishable, whereas as the temperature becomes a sizable fraction of the Planck 
temperature $T_{\rm P}$ the absence of photon states with $E > E_{\rm P}$ becomes thermodynamically 
relevant. The equation-of-state ratio $P/u$ approaches the SR value $1/3$ in the low-temperature limit, 
with the leading correction exponentially suppressed as $\sim x_{\rm P}^3 e^{-x_{\rm P}}$, and grows 
monotonically thereafter; as we have shown in Sec.~\ref{sec:eos}, in the formal limit 
$x_{\rm P}\to 0$ it diverges logarithmically while the sound speed grows without bound. However, such a 
causality violation is based on the standard relativistic isentropic sound-speed formula without
justification within the deformed boost framework, is interpretive rather than definitive for DSR-type 
constructions.

The low-temperature results are robust against the choice of measure: for generic nonzero $\theta$, 
expanding the partition function about the SR result modifies the corrections of the flat-measure case 
only through the factor 
$(1-E/E_{\rm P})^{-\theta}$ evaluated near $E\approx k_{\rm B} T$, i.e.\ only at relative order 
$1/x_{\rm P}$ and higher. The recovery of the SR limit, including in particular $P/u=1/3$, is 
therefore independent of the measure choice; the character of the leading correction is not. 
For $\theta=0$, the deviations from SR are exponentially suppressed, 
$\sim e^{-x_{\rm P}}$; for any nonzero admissible $\theta$, they are power-law suppressed, 
$\sim\theta/x_{\rm P}$. By contrast, every statement 
at $T\gtrsim E_{\rm P}$ --- the suppression pattern, the saturation of the heat capacity and the 
modified equation of state --- originates in the endpoint behaviour of the truncated integral and is an 
artefact of the hard-cutoff prescription rather than a property of MS kinematics.
We therefore restrict the physical interpretation of our 
results to $T\lesssim E_{\rm P}$. Our calculation also clarifies and corrects an inconsistency present 
in some earlier treatments of photon gas thermodynamics in DSR.

Finally, the model dependence of these conclusions deserves to be made explicit. In the Amelino-Camelia 
realization the photon dispersion relation is genuinely deformed, Eq.~\eqref{eq:ACdisp}, so that the 
density of states is modified at all energies and the deviations from SR are controlled by $E/\kappa$ 
($\kappa$ denotes the invariant energy scale) rather than by $e^{-E_{\rm P}/k_{\rm B}T}$; a complete 
parallel thermodynamic treatment requires a 
separate calculation and we will address this problem in a future publication. 

\section{Acknowledgement}
The authors would like to thank the referee for insightful comments which greatly improved 
the quality of the paper. This work is partially supported by the National Natural Science 
Foundation of China (NSFC) via Grant No. 11505115.



\begin{thebibliography}{99}

\bibitem{Ameli2001a}
G. Amelino-Camelia and T. Piran, Phys. Rev. D \textbf{64}, 036005 (2001).

\bibitem{Ameli2001b}
G. Amelino-Camelia, Phys. Lett. B \textbf{510}, 255 (2001).

\bibitem{Glik2001}
J. Kowalski-Glikman, Phys. Lett. A \textbf{286}, 391 (2001).

\bibitem{Ameli2002a}
G. Amelino-Camelia, Nature \textbf{418}, 34 (2002).

\bibitem{Ameli2002b}
G. Amelino-Camelia, Int. J. Mod. Phys. D \textbf{11}, 35 (2002).

\bibitem{Mague2002}
J. Magueijo and L. Smolin, Phys. Rev. Lett. \textbf{88}, 190403 (2002).

\bibitem{Glik2002a}
J. Kowalski-Glikman, Phys. Lett. A \textbf{299}, 454 (2002).

\bibitem{Glik2002b}
J. Kowalski-Glikman and S. Nowak, Phys. Lett. B \textbf{539}, 126 (2002).

\bibitem{Mague2003}
J. Magueijo and L. Smolin, Phys. Rev. D \textbf{67}, 044017 (2003).

\bibitem{Glik2003}
J. Kowalski-Glikman and S. Nowak, Class. Quant. Grav. \textbf{20}, 4799 (2003).

\bibitem{Mague2004}
J. Magueijo and L. Smolin, Class. Quant. Grav. \textbf{21}, 1725 (2004).

\bibitem{Glik2005}
J. Kowalski-Glikman, Lect. Notes Phys. \textbf{669}, 131 (2005).

\bibitem{Ameli2010a}
G. Amelino-Camelia, Symmetry \textbf{2}, 230 (2010).

\bibitem{Ameli2000}
G. Amelino-Camelia and S. Majid, Int. J. Mod. Phys. A \textbf{15} (2000) 4301.

\bibitem{Ameli2009}
G. Amelino-Camelia, N. Loret, G. Mandanici and F. Mercati, arXiv:0906.2016.

\bibitem{Chan2012}
N. Chandra and S. Chatterjee, Phys. Rev. D \textbf{85}, 045012 (2012).

\bibitem{Sales2009}
G. Salesi and E. Di Grezia, Phys. Rev. D \textbf{79}, 104009 (2009).

\bibitem{Ameli2006}
G. Amelino-Camelia, M. Arzano, Y. Ling, and G. Mandanici, Class. Quant. Grav. \textbf{23}, 2585 (2006).

\bibitem{Alex2003}
S. Alexander, R. H. Brandenberger, and J. Magueijo, Phys. Rev. D \textbf{67}, 081301 (2003).

\bibitem{Greg2009}
M. Gregg and S. A. Major, Int. J. Mod. Phys. D \textbf{18}, 971 (2009).

\bibitem{Berto2010}
O. Bertolami and C. A. D. Zarro, Phys. Rev. D \textbf{81}, 025005 (2010).

\bibitem{Gamb1999}
R. Gambini and J. Pullin, Phys. Rev. D \textbf{59}, 124021 (1999).

\bibitem{Ellis2000}
J. Ellis,  K. Farakos, N. E. Mavromatos, V. Mitsou, D. V. Nanopoulos, Astrophys. J. \textbf{535}, 139 (2000).

\bibitem{Alfaro1999}
J. Alfaro, H. A. Morales-Tecotl, and L. F. Urrutia, Phys. Rev. Lett. \textbf{84}, 2318 (2000).

\bibitem{Mersini2001}
L. Mersini, M. Basterogil, and P. Kanti, Phys. Rev. D \textbf{64}, 043508 (2001).

\bibitem{Menc2013}
L. Menculini, O. Panella, and P. Roy, Phys. Rev. D \textbf{87}, 065017 (2013).

\bibitem{Jafa2024}
N. Jafari, B. Shukirgaliyev, Phys. Lett. B \textbf{853}, 138693 (2024).

\bibitem{Guven2025}
A. Guvendi and O. Mustafa, Eur. Phys. J. C \textbf{85}, 1027 (2025).

\bibitem{Ghosh2007}
S. Ghosh and P. Pal, Phys. Rev. D \textbf{75}, 105021 (2007).

\bibitem{Bruno2001} 
N. R. Bruno, G. Amelino-Camelia and J. Kowalski-Glikman, Phys. Lett. B \textbf{522}, 133 (2001). 

\bibitem{Ameli2010b} 
G. Amelino-Camelia, Recent Developments in Theoretical Physics 
(Statistical Science and Interdisciplinary Research - Vol. 9),
S. Ghosh and G. Kar, ed. World Scientific, Singapore (2010).

\bibitem{Ameli2013}
G. Amelino-Camelia, Living Reviews in Relativity \textbf{16}, 5 (2013).

\bibitem{Carm2019}
J. M. Carmona, J. L. Cortes, and J. J. Relancio, Symmetry 11, 1401 (2019).

\bibitem{Carm2023}
J. M. Carmona,J. L. Cort\'{e}s,J. J. Relancio, and M. A. Reyes, Universe 9(3), 150 (2023).

\bibitem{Hoss2006} 
S. Hossenfelder, Class. Quant. Grav. \textbf{23}, 1815 (2006). 

\bibitem{Cama2007}
A. Camacho and A. Macias, Gen. Rel. Grav. \textbf{39}, 1175 (2007).

\bibitem{Das2010}
S. Das and D. Roychowdhury, Phys. Rev. D \textbf{81}, 085039 (2010).

\bibitem{KH1987}
K. Huang, Statistical Mechanics, 2nd. ed., John Wiley \& Sons (1987).

\bibitem{Gubito2016}
G. Gubitosi and J. Magueijo, Class. Quant. Grav. \textbf{33}, 115021 (2016). 

\bibitem{Rembie2003}
J. Rembieli\'nski and K. A. Smoli\'nski, Bull. Soc. Sci. Lett. Lodz \textbf{53}, 57 (2003).

\bibitem{Gorji2016}
M. A. Gorji, V. Hosseinzadeh, K. Nozari, and B. Vakili, J. Stat. Mech. (JSTAT) (2016) 073107.

\bibitem{Blaut2004}
A. B\l aut, M. Daszkiewicz, J. Kowalski-Glikman, and S. Nowak, Phys. Lett. B \textbf{582}, 82 (2004).

\bibitem{Zhang2011}
X. Zhang, L. Shao, and B.-Q. Ma, Astropart. Phys. \textbf{34}, 840 (2011).

\bibitem{Jafa2025}
N. Jafari and A. Boumali, Phys. Lett. B \textbf{871}, 140019 (2025).

\bibitem{Abra1972}
M. Abramowitz and I. A. Stegun, Handbook of Mathematical Functions with Formulas,
Graphs, and Mathematical Tables, New York: Dover Publications (1972).

\end{thebibliography}
\end{document}